# Diffusion and Correlations in Lattice Gas Automata

David Hanon and Jean Pierre Boon *
*Center for Nonlinear Phenomena and Complex Systems*
*Université Libre de Bruxelles, Campus Plaine, C.P. 231*
*1050 Bruxelles, Belgium*
(November 10, 2018)

We present an analysis of diffusion in terms of the spontaneous density fluctuations in a non-thermal two-species fluid modeled by a lattice gas automaton. The power spectrum of the density correlation function is computed with statistical mechanical methods, analytically in the hydrodynamic limit, and numerically from a Boltzmann expression for shorter time and space scales. In particular we define an observable – the weighted difference of the species densities – whose fluctuation correlations yield the diffusive mode independently of the other modes so that the corresponding power spectrum provides a measure of diffusion dynamics solely. Automaton simulations are performed to obtain measurements of the spectral density over the complete range of wavelengths (from the microscopic scale to the macroscopic scale of the automaton universe). Comparison of the theoretical results with the numerical experiments data yields the following results: (i) the spectral functions of the lattice gas fluctuations are in accordance with those of a classical 'non-thermal' fluid; (ii) the Landau-Placzek theory, obtained as the hydrodynamic limit of the Boltzmann theory, describes the spectra correctly in the long wavelength limit; (iii) at shorter wavelengths and at moderate densities the complete Boltzmann theory provides good agreement with the simulation data. These results offer convincing validation of lattice gas automata as a microscopic approach to diffusion phenomena in fluid systems.

PACS numbers: 05.20.Dd, 05.50.+q, 05.60.+w

## I. INTRODUCTION

Frisch, Hasslacher and Pomeau pioneered a lattice gas automaton (FHP) as a microscopic model for incompressible fluids obeying the Navier-Stokes equation in two dimensions [1]. The FHP model was subsequently generalized to study diffusive phenomena in binary fluids using 'macroscopic' experiments [2–4]. Typically the observer would be interested in the evolution of the density profile of 'red' particles in a system composed of 'red' and 'blue' particles where the color is a passive property used to distinguish species which otherwise do not differ from one another (it is necessary that they do in other circumstances [5]). The diffusion coefficient is then evaluated by fitting the 'experimental' profile to the solution of the diffusion equation subject to the appropriate boundary conditions [3,4] .

Because the FHP lattice gas lacks an independent collisional invariant for energy, it is not suited for the modeling of thermal fluids. An appropriate generalization was realized by the construction of the model proposed by Grosfils, Boon and Lallemand (GBL) [6]. Their study was motivated by the analysis of the correlations of spontaneous fluctuations in lattice gas automata (LGA) in order to find whether the fluctuations power spectrum would be in accordance with those observed in actual fluids. Indeed the dynamical structure factor – the power spectrum of the density fluctuations correlation function – gains its importance by providing insight to the dynamical behavior of the fluid [7], and the LGA was found to exhibit correct properties at global equilibrium: the spectra obtained by simulations of the GBL model present the same characteristics as those obtained from neutron- and light-scattering experiments in real fluids. In particular, in the hydrodynamic limit, one observes two shifted Brillouin peaks (corresponding to the propagation of sound waves and their damping) and a central Rayleigh peak (corresponding to the diffusivity of entropy fluctuations as a consequence of energy conservation). The GBL model was subsequently analyzed in detail by Grosfils *et al* [8].

The mixture of two real fluids exhibits a power spectrum in which the central peak is not a simple Lorentzian, even in the long wavelength limit [7]. It has a spectral structure where it is difficult to separate the contributions from entropy fluctuations and from concentration fluctuations which are not decoupled in general (unless one of the two components is in trace amounts, in which case the two modes can be identified as they produce two independent central Lorentzians).

*E-mail addresses: `david.hanon@ulb.ac.be, jpboon@ulb.ac.be`

From the above considerations the idea emerged to analyze and measure the fluctuation correlations in a non-thermal two-species LGA fluid (in which the Rayleigh peak is absent) in order to study diffusion dynamics from a microscopic approach. In section II we present the model used for our numerical simulations. Section III discusses the lattice Boltzmann theory for the analysis of the dynamical structure factor. The analytical results are developed in the hydrodynamic limit in section IV and are found to be in full agreement with the Landau-Placzek theory. In section V we examine the different wavelength domains in terms of the Boltzmann propagator eigenvalues, and we present qualitative and quantitative analyzes of the theoretical results in comparison with the simulation data. The FHP spurious invariant is identified and its effects are shown to be unimportant when the LGA is properly implemented. We conclude with some comments.

## II. THE MODEL

The particles have unitary mass with no spatial extension and occupy the nodes of a triangular lattice with hexagonal symmetry. A particle can move along any of the six lattice directions (with unit velocity modulus) to one of the nearest sites or be at rest in its initial state (with zero velocity modulus). Particles interact via instantaneous local collisions which redistribute mass and momentum among the channels of each node at every time-step according to mass and momentum conservation. In a two-species system, particles are tagged either as 'red' or 'blue', and their color is redistributed randomly during the collisions independently of the mass redistribution; color is also conserved by the dynamics. The state of a node is given in terms of channel occupations. Here we use a description assigning a color to the channel. Since there are seven distinct velocities and two colors (one for each species), each node has seven pairs of channels. An exclusion principle is applied such that a pair of channels cannot be occupied by more than one particle (either red or blue) at any given time [9]. A corollary is that the equilibrium distribution takes the form of a Fermi-Dirac distribution [10]. The present formulation yields a convenient specification of the state of a node as a fourteen-bit word.

## III. THE BOLTZMANN FORMALISM

The red mass density $\rho^{red}(\mathbf{r}, \mathbf{t})$ is the number of red particles at node $\mathbf{r}$ at time $t$ and the fluctuations $\delta\rho^{red}(\mathbf{r}, t)$ are defined in terms of the red channel occupations $n_i$ ($i \in \{1, ..., b\}$, with $2b$ the total number of channels per node)

$$\delta\rho^{red}(\mathbf{r}, t) \;=\; \sum_{i=1}^{b} \delta n_i(\mathbf{r}, t) \;=\; \sum_{i=1}^{b} [n_i(\mathbf{r}, t) - < n_i(\mathbf{r}, t) >] \,, \tag{1}$$

where $<>$ denotes the equilibrium ensemble average; in basic equilibrium

$$< n_i(\mathbf{r}, t) > \equiv f_i = \begin{cases} f\theta & \text{for } i = 1, \ldots, b \text{ (red channels)}, \\ f(1-\theta) & \text{for } i = b+1, \ldots, 2b \text{ (blue channels)}, \end{cases} \tag{2}$$

with $f$ the average density per pair of channels, and $\theta$ the concentration of red particles. Note that

$$\sum_{i=1}^{2b} < n_i(\mathbf{r}, t) > = bf\theta + bf(1-\theta) = \rho^{red} + \rho^{blue} = \rho \,, \tag{3}$$

which defines the respective average densities per node.

The 'red mass' dynamic structure factor $S^{red}(\mathbf{k}, \omega)$, defined as the space and time Fourier transform of the van Hove correlation function

$$G^{red}(\mathbf{r}, t) = < \delta\rho^{red}(\mathbf{r}, t) \; \delta\rho^{red}(0, 0) > \,, \tag{4}$$

is given by

$$\rho^{red}\, S^{red}(\mathbf{k},\omega) = \sum_{\mathbf{r}\in\mathcal{L}} \sum_{t=-\infty}^{\infty} e^{-i\omega t - i\mathbf{k}\cdot\mathbf{r}} G^{red}(\mathbf{r},|t|) \tag{5}$$

$$= \frac{1}{V} \sum_{t=-\infty}^{\infty} e^{-i\omega t} < \delta\rho^{red}(\mathbf{k},|t|)\, \delta\rho^{red\,*}(\mathbf{k},0) >, \tag{6}$$

where $\delta\rho^{red}(\mathbf{k},t)$ is the spatial Fourier transform of $\delta\rho^{red}(\mathbf{r},t)$, and $V$ is the total number of nodes, also interpreted as the volume of the lattice universe (here the lattice $\mathcal{L}$ is finite and has periodic boundary conditions). $S^{red}(\mathbf{k},\omega)$ is also expressed in terms of the kinetic propagator [8] defined by

$$\Gamma_{ij}(\mathbf{k},t)\; \kappa_j = < \delta n_i(\mathbf{k},t)\, \delta n_j^*(\mathbf{k},0) > , \quad i,j \;=\; 1,...,b\, , \tag{7}$$

where $\delta n_i(\mathbf{k},t)$ is the spatial Fourier transform of $\delta n_i(\mathbf{r},t)$, and $\kappa_j = f_j(1-f_j)$. Using (7), we write the dynamic structure factor (5) as

$$\rho^{red}\, S^{red}(\mathbf{k},\omega) = \sum_{t=-\infty}^{\infty} e^{-i\omega t} \sum_{i=1}^{b}\sum_{j=1}^{b} \Gamma_{ij}(\mathbf{k},t)\; \kappa_j\, , \tag{8}$$

and the static structure factor (the Fourier transform of the equal-time van Hove function) as

$$\rho^{red}\, S^{red}(\mathbf{k}) = \sum_{i=1}^{b}\sum_{j=1}^{b} \Gamma_{ij}(\mathbf{k},0)\; \kappa_j \tag{9}$$

$$= \sum_{i=1}^{b}\sum_{j=1}^{b} \delta_{ij}\; \kappa_j = \sum_{j=1}^{b} \kappa_j\, , \tag{10}$$

or

$$S^{red}(\mathbf{k}) = 1 - f\theta\, . \tag{11}$$

We now evaluate the kinetic propagator in the Boltzmann approximation [11]. The lattice gas equation for the *single particle distribution* $f_i(\mathbf{r},t)$ reads [12]

$$f_i(\mathbf{r}+\mathbf{c}_i, t+1) = f_i(\mathbf{r},t) + \Delta(\{n_j\}). \tag{12}$$

Here $\Delta(\{n_i\})$ is the collision term, which is expanded around the stationary equilibrium distribution $< n_i >$ to yield

$$\Delta(\{< n_i > + \delta n_i\}) = \sum_{j=1}^{2b} \Omega_{ij}\delta n_j + \sum_{j,k} \mathcal{O}(\delta n_j \delta n_k), \tag{13}$$

where we have used the property $\Delta(\{< n_i >\}) = 0$ which follows from mass conservation. The explicit form of $\Omega_{ij}$ is given in terms of the transition matrix $A(s \to s')$ between pre- and post-collisional states $s$ and $s'$ respectively

$$\Omega_{ij} = \sum_{\{s,s'\}} A(s \to s')(s'_i - s_i)s_j \prod_{k=1}^{b} \frac{< n_k >^{s_k} < 1 - n_k >^{1-s_k}}{< n_j >< 1 - n_j >}. \tag{14}$$

This result is obtained with the assumption that particles on different channels of the same node are uncorrelated before collision (Boltzmann *ansatz*), i.e. by factorizing the averages $< n_i\, n_j >$ $(i \neq j)$. Combining (12) and (13), we obtain the linearized Boltzmann equation which reads in $\mathbf{k}$-space

$$\delta n_i(\mathbf{k}, t+1) = \sum_{j=1}^{2b} e^{-i\mathbf{k}\cdot\mathbf{c}_i}\, (\delta_{ij} + \Omega_{ij})\, \delta n_j(\mathbf{k},t). \tag{15}$$

Equation (15) is straightforwardly solved by iteration; inserting its solution into (7) yields

$$\Gamma_{ij}(\mathbf{k},t)\, \kappa_j = \left[\mathbf{e}^{-i\mathbf{k}\cdot\mathbf{c}} \cdot (\boldsymbol{\delta} + \boldsymbol{\Omega})\right]^t_{ij} \kappa_j\; , \; (t \geq 0). \tag{16}$$

Here $[\mathbf{e}^{-i\mathbf{k}\cdot\mathbf{c}}]_{j\ell} = \delta_{j\ell} e^{-i\mathbf{k}\cdot\mathbf{c}_j}$ is a diagonal matrix. From (8) and (16) we obtain

$$\rho^{red}\, S^{red}(\mathbf{k},\omega) \equiv 2\,\mathcal{R}e F^{red}(\mathbf{k},\omega), \tag{17}$$

$$F^{red}(\mathbf{k},\omega) = \sum_{i=1}^{b}\sum_{j=1}^{b}\left[\frac{1}{\mathbf{e}^{i\omega+i\mathbf{k}\cdot\mathbf{c}} - \boldsymbol{\delta} + \boldsymbol{\Omega}} + \frac{1}{2}\right]_{ij} \kappa_j\,, \tag{18}$$

where $\mathcal{R}e$ denotes the real part. This expression for the dynamic structure factor is exact within the Boltzmann approximation, but the explicit analytical inversion of the $b \times b$ matrix in (18) cannot be performed in all generality. However perturbation methods can be used to compute analytically $S^{red}(\mathbf{k},\omega)$ in the hydrodynamic limit: $|\mathbf{k}| \longrightarrow 0$ and $\omega \sim \mathcal{O}(|\mathbf{k}|)$, $\mathcal{O}(|\mathbf{k}|^2)$ (Section IV). Beyond the long wavelength – long time domain, one has recourse to numerical evaluation of (18) to compute the 'Boltzmann' power spectrum (Section V).

## IV. THE HYDRODYNAMIC LIMIT

### A. The hydrodynamic modes

We first notice that the linearized collision operator $\boldsymbol{\Omega}$ is not symmetrical, with the consequence that its left and right eigenvectors are not each others transpose. However when the detailed balance condition is satisfied, the matrix product $\Omega_{ij}\kappa_j$ is symmetrical [13], and the left and right eigenvectors of $\boldsymbol{\Omega}$ are related by $|\phi\rangle_i = \kappa_i\ \langle\phi|_i$; it can also be shown that to each of the $N$ collisional invariants corresponds an eigenvector $\langle A_n|(n = 1,...,N)$ of $\boldsymbol{\Omega}$, with zero eigenvalue. The components are given by the conserved quantities carried by each channel $i$ :

$$\begin{array}{lll} \text{red mass} & : & \langle R|_i = 1\,, \quad \text{if } i = 1,...,b\,;\ \ \langle R|_i = 0\,, \quad \text{if } i = b+1,...,2b\,; \\ \text{blue mass} & : & \langle B|_i = 0\,, \quad \text{if } i = 1,...,b\,;\ \ \langle B|_i = 1\,, \quad \text{if } i = b+1,...,2b\,; \\ x\text{-momentum} & : & \langle P_x|_i = \mathbf{c}_i \cdot \mathbf{1}_x\,; \\ y\text{-momentum} & : & \langle P_y|_i = \mathbf{c}_i \cdot \mathbf{1}_y\,. \end{array} \tag{19}$$

From these considerations and by analogy with the *thermal* scalar product introduced in [8], we define the *colored* scalar product

$$\langle A|B\rangle = \sum_{i=1}^{b} A(\mathbf{c}_i)\,\kappa_i\,B(\mathbf{c}_i)\,, \tag{20}$$

where the weight $\kappa_i$ depends on density and concentration. Since $\Omega_{ij}\kappa_j$ is a symmetrical matrix, the colored scalar product has the symmetry

$$\langle A|\boldsymbol{\Omega}|B\rangle \ = \ \langle B|\boldsymbol{\Omega}|A\rangle \ = \ \sum_{i=1}^{2b}\sum_{j=1}^{2b} A(\mathbf{c}_i)\,\Omega_{ij}\kappa_j\,B(\mathbf{c}_j)\,. \tag{21}$$

Following a method introduced by Résibois [14], we consider, as the starting point, the propagator (16) which is the $t$-th power of the non-symmetrical matrix $\mathbf{e}^{-i\mathbf{k}\cdot\mathbf{c}} \cdot (\boldsymbol{\delta} + \boldsymbol{\Omega})$, and we use the eigenvalue problem formulation

$$\mathbf{e}^{-i\mathbf{k}\cdot\mathbf{c}} \cdot (\boldsymbol{\delta} + \boldsymbol{\Omega})\ |\psi_\mu(\mathbf{k})\rangle = e^{z_\mu(\mathbf{k})}\ |\psi_\mu(\mathbf{k})\rangle\,, \tag{22}$$

$$\langle\phi_\mu(\mathbf{k})|\ \mathbf{e}^{-i\mathbf{k}\cdot\mathbf{c}} \cdot (\boldsymbol{\delta} + \boldsymbol{\Omega}) = e^{z_\mu(\mathbf{k})}\ \langle\phi_\mu(\mathbf{k})|\,. \tag{23}$$

The eigenmodes of the propagator may be separated into two groups: the slow modes, corresponding to eigenvalues $z_\mu(\mathbf{k})$ close to zero when $k(= |\mathbf{k}|)$ tends to zero, and the fast modes corresponding to eigenvalues $\mathcal{R}e\, z_\mu(\mathbf{k}) < 0$ leading to exponentially fast decay. The latter are the kinetic modes; the slow modes which decay infinitely slowly when $k \to 0$ will be identified as the hydrodynamic modes. They are the dominant modes in the hydrodynamic regime where the kinetic modes can be neglected.

For $|\mathbf{k}| = 0$ the matrix $\mathbf{e}^{-i\mathbf{k}\cdot\mathbf{c}} \cdot (\boldsymbol{\delta} + \boldsymbol{\Omega})$ reduces to $\boldsymbol{\delta} + \boldsymbol{\Omega}$, whose eigenspace spanned by the eigenvectors (19) has the dimension given by the number of collisional invariants (here 4). For $|\mathbf{k}| \neq 0$ but small, we can express the eigenvectors of $\mathbf{e}^{-i\mathbf{k}\cdot\mathbf{c}} \cdot (\boldsymbol{\delta} + \boldsymbol{\Omega})$ as a linear combination of the collisional invariants, and we can expand $e^{-i\mathbf{k}\cdot\mathbf{c}}$, $|\psi_\mu(\mathbf{k})\rangle$, $\langle\phi_\mu(\mathbf{k})|$, and $z_\mu(\mathbf{k})$ respectively as

$$
\begin{array}{rcllllll}
\mathbf{e}^{-i\mathbf{k}\cdot\mathbf{c}} & = & \boldsymbol{\delta} & - & (ik)\,\mathbf{c}_\ell & + & \frac{1}{2}(ik)^2\,\mathbf{c}_\ell^2 & -\dots\,, \\
|\psi_\mu(\mathbf{k})\rangle & = & |\psi_\mu^{(0)}\rangle| & + & (ik)\,|\psi_\mu^{(1)}\rangle & + & (ik)^2\,|\psi_\mu^{(2)}\rangle & +\dots\,, \\
\langle\phi_\mu(\mathbf{k})| & = & \langle\phi_\mu^{(0)}| & + & (ik)\,\langle\phi_\mu^{(1)}| & + & (ik)^2\,\langle\phi_\mu^{(2)}| & +\dots\,, \\
z_\mu(\mathbf{k}) & = & & & (ik)\,z_\mu^{(1)} & + & (ik)^2\,z_\mu^{(2)} & +\dots\,,
\end{array}
\tag{24}
$$

with $c_{\ell,ij} = \delta_{ij}\,c_{\ell,i}$; $i,j = 1,...,2b$, where $c_{\ell,i}$ denotes the projection of $\mathbf{c}_i$ onto $\mathbf{k}$. Substitution of the first and second expressions of (24) into (22) and identification of the successive powers of $k$ yields the hierarchy

$$
\mathcal{O}(k^0):\quad \boldsymbol{\Omega}|\psi_\mu^{(\mathbf{0})}\rangle = 0\,, \tag{25}
$$

$$
\mathcal{O}(k^1):\quad \boldsymbol{\Omega}|\psi_\mu^{(\mathbf{1})}\rangle = \left(\mathbf{c}_\ell + z_\mu^{(1)}\boldsymbol{\delta}\right)|\psi_\mu^{(0)}\rangle\,, \tag{26}
$$

$$
\mathcal{O}(k^2):\quad \boldsymbol{\Omega}|\psi_\mu^{(\mathbf{2})}\rangle = \left(\mathbf{c}_\ell + z_\mu^{(1)}\boldsymbol{\delta}\right)|\psi_\mu^{(1)}\rangle + \left[z_\mu^{(2)}\boldsymbol{\delta} + \frac{1}{2}\left(\mathbf{c}_\ell + z_\mu^{(1)}\boldsymbol{\delta}\right)^2\right]|\psi_\mu^{(0)}\rangle\,. \tag{27}
$$

The solution to zero-th order is straightforward; one has

$$
|\psi_\mu^{(0)}\rangle = \sum_{n=1}^{N} b_n|A_n\rangle\,, \tag{28}
$$

where the coefficients $b_n$ are to be determined subsequently.

The first order solution is obtained by taking the scalar product of $\langle A_m|$ with (26) where the previous order solution is substituted; the result has the form as an $N$-dimensional eigenvalue problem:

$$
\sum_{n=1}^{N}\langle A_m|\left(\mathbf{c}_\ell + z_\mu^{(1)}\boldsymbol{\delta}\right)|A_n\rangle\,b_n\,,\,=0\,, \tag{29}
$$

which yields the four eigenvectors $|\psi_\mu^{(0)}\rangle$ and eigenvalues $z_\mu^{(1)}$

$$
\begin{array}{llcllcl}
\text{shear mode:} & |\psi_\perp^{(0)}\rangle & = & |P_\perp\rangle & ; & z_\perp^{(1)} = & 0\,, \\
\text{acoustic modes:} & |\psi_+^{(0)}\rangle & = & |P_\ell\rangle - c_s|M\rangle & ; & z_+^{(1)} = & +c_s\,, \\
 & |\psi_-^{(0)}\rangle & = & |P_\ell\rangle + c_s|M\rangle & ; & z_-^{(1)} = & -c_s\,, \\
\text{color diffusion mode:} & |\psi_{\text{diff}}^{(0)}\rangle & = & \kappa_b|R\rangle - \kappa_r|B\rangle & ; & z_{\text{diff}}^{(1)} = & 0\,.
\end{array}
\tag{30}
$$

Here $|M\rangle$ is the sum of $|R\rangle$ and $|B\rangle$, $P_\ell$ and $P_\perp$ are the projections of the momentum onto $\mathbf{k}$ and perpendicular to $\mathbf{k}$ respectively, $\kappa_j = f_j(1-f_j)$ with $j = 1,...,b$ for $\kappa_r$, and $j = b+1,...,2b$ for $\kappa_b$, and $c_s = (\langle P_\ell|P_\ell\rangle/\langle M|M\rangle)^{1/2}$ will be identified as the speed of sound (here $c_s = \sqrt{3/7}$).

We define the currents $|j_\mu\rangle$ as

$$
|j_\mu\rangle = \left(\mathbf{c}_\ell + z_\mu^{(1)}\boldsymbol{\delta}\right)|\psi_\mu^{(0)}\rangle\,, \tag{31}
$$

and we note, by multiplication of (26) by $\langle\psi_\nu|^{(0)}$, that the currents are orthogonal to $\langle\psi_\nu^{(0)}|$. As a consequence the currents do not belong to the $\boldsymbol{\Omega}$ kernel, and we may write the formal solution to the first order equation (26) as

$$
|\psi_\mu^{(1)}\rangle = \frac{1}{\boldsymbol{\Omega}}|j_\mu\rangle + \sum_{\nu=1}^{N} b_{\mu\nu}|\psi_\nu^{(0)}\rangle \tag{32}
$$

The coefficients $b_{\mu\nu}$ are determined by substituting $|\psi_\mu^{(1)}\rangle$ by (32) in the second order equation (27), and multiplying the result by $\langle\psi_\nu^{(0)}|$ $(\nu \neq \mu)$ to obtain

$$
b_{\mu\nu}\langle\psi_\nu^{(0)}|\psi_\nu^{(0)}\rangle(z_\mu^{(1)} - z_\nu^{(1)}) = -\langle j_\nu|\frac{1}{\boldsymbol{\Omega}} + \frac{\boldsymbol{\delta}}{2}|j_\mu\rangle. \tag{33}
$$

The expression for $z_\mu^{(2)}$ follows from the evaluation of the product of equation (27) with $\langle\psi_\mu^{(0)}|$, which yields

$$
z_\mu^{(2)} = \frac{\langle j_\mu|\left(\frac{1}{\boldsymbol{\Omega}} + \frac{\boldsymbol{\delta}}{2}\right)|j_\mu\rangle}{\langle\psi_\mu^{(0)}|\psi_\mu^{(0)}\rangle}\,. \tag{34}
$$

We anticipate that $z_\ell^{(2)}$ is the kinematic viscosity ($\nu$), that $z_+^{(2)} = z_-^{(2)}$ is the sound damping ($\Gamma$), and that $z_{\rm diff}^{(2)}$ is the color-diffusivity ($D$), as will be justified subsequently by the analysis of the power spectrum.

Explicit evaluation of (33) shows that the only non-zero off-diagonal elements of the matrix formed by the $b_{\mu\nu}$'s are the two coefficients

$$b_{+,-} \;=\; -\; b_{-,+} = \;\frac{\Gamma}{2\,c_s}. \tag{35}$$

The diagonal elements $b_{\mu\mu}$ remain unknown, but this is unimportant because, as will be seen, they do not contribute to the power spectrum (Section V).

We have now identified the four hydrodynamic modes in the LGA. The shear mode and the acoustic modes are independent of color related properties, and the mode $|\psi_{\rm diff}\rangle$ describes color diffusion only. As will be shown below, the density power spectrum reflects this property. Notice that the purely diffusive behavior of color is related to an observable (defined below) which is neither the concentration of one of the components nor the difference between the two concentrations [3,4].

### B. The dynamic structure factor

Transposing Eq.(23) which defines the left eigenvectors of $\mathbf{e}^{-i\mathbf{k}\cdot\mathbf{c}}\cdot(\boldsymbol{\delta}+\boldsymbol{\Omega})$, and multiplying the result by $\mathbf{e}^{-i\mathbf{k}\cdot\mathbf{c}}$ on the left, we find that $\langle\phi_\mu|$ and $|\psi_\mu\rangle$ are related by

$$|\phi_\mu(\mathbf{k})\rangle = \frac{1}{m_\mu}\,\mathbf{e}^{+i\mathbf{k}\cdot\mathbf{c}}|\psi_\mu(\mathbf{k})\rangle\,, \tag{36}$$

where $m_\mu$ is a normalization constant. If the eigenvectors $\langle\phi_\mu|$ and $|\psi_\mu\rangle$ form a complete bi-orthonormal set, i.e.

$$\sum_{\mu=1}^{2b}|\psi_\mu(\mathbf{k})\rangle\langle\phi_\mu(\mathbf{k})| = \boldsymbol{\delta}\;,\;\;\text{and}\;\;\langle\phi_\mu(\mathbf{k})|\psi_\nu(\mathbf{k})\rangle = \delta_{\mu\nu}\,, \tag{37}$$

we may write $\mathbf{e}^{-i\mathbf{k}\cdot\mathbf{c}}\cdot(\boldsymbol{\delta}+\boldsymbol{\Omega})$ as

$$\mathbf{e}^{-i\mathbf{k}\cdot\mathbf{c}}\cdot(\boldsymbol{\delta}+\boldsymbol{\Omega}) = \sum_{\mu=1}^{2b}|\psi_\mu(\mathbf{k})\rangle e^{z_\mu(\mathbf{k})}\langle\phi_\mu(\mathbf{k})|\,. \tag{38}$$

We will use this expression to recast the spectral function $F^{red}(\mathbf{k},\omega)$ ; we first rewrite (18) as

$$\begin{aligned}
F^{red}(\mathbf{k},\omega) &= \sum_{i=1}^{b}\sum_{j=1}^{b}\left[\frac{1}{\mathbf{e}^{i\omega+i\mathbf{k}\cdot\mathbf{c}}-\boldsymbol{\delta}-\boldsymbol{\Omega}}+\frac{1}{2}\boldsymbol{\delta}\right]_{ij}\kappa_j\\
&= \langle R|\frac{1}{\mathbf{e}^{i\omega+i\mathbf{k}\cdot\mathbf{c}}-\boldsymbol{\delta}-\boldsymbol{\Omega}}+\frac{1}{2}\boldsymbol{\delta}|R\rangle\\
&= \langle R|\frac{1}{e^{i\omega}\left(\mathbf{e}^{-i\mathbf{k}\cdot\mathbf{c}}\cdot(\boldsymbol{\delta}+\boldsymbol{\Omega})\right)^{-1}-\boldsymbol{\delta}}\cdot(\boldsymbol{\delta}+\boldsymbol{\Omega})^{-1}+\frac{1}{2}\boldsymbol{\delta}|R\rangle\\
&= \langle R|\frac{1}{e^{i\omega}\left(\mathbf{e}^{-i\mathbf{k}\cdot\mathbf{c}}\cdot(\boldsymbol{\delta}+\boldsymbol{\Omega})\right)^{-1}-\boldsymbol{\delta}}+\frac{1}{2}\boldsymbol{\delta}|R\rangle\,,
\end{aligned} \tag{39}$$

where the last equality is obtained by noticing that the action of $(\boldsymbol{\delta}+\boldsymbol{\Omega})^{-1}$ upon $|R\rangle$ is the identity operation since $|R\rangle$ belongs to the kernel of $\boldsymbol{\Omega}$. Then making use of (38), we find

$$\begin{aligned}
F^{red}(\mathbf{k},\omega) &= \sum_{\mu=1}^{b}\langle R|\psi_\mu(\mathbf{k})\rangle\left[\frac{1}{e^{i\omega-z_\mu(\mathbf{k})}-1}+\frac{1}{2}\right]\langle\phi_\mu(\mathbf{k})|R\rangle \;=\; \sum_{\mu=1}^{b}\mathcal{N}_\mu\mathcal{D}_\mu\,,\\
\text{with}\quad \mathcal{N}_\mu &= \langle R|\psi_\mu(\mathbf{k})\rangle\langle\phi_\mu(\mathbf{k})|R\rangle\,,\quad\text{and}\quad \mathcal{D}_\mu = \frac{1}{e^{i\omega-z_\mu(\mathbf{k})}-1}+\frac{1}{2}\,.
\end{aligned} \tag{40}$$

We observe that each mode $\mu$ contributes a spectral line whose amplitude depends on $\mathcal{D}_\mu$. This factor becomes large for $(i\omega - z_\mu)\to 0$, that is for small $z_\mu$ in the limit of small $\omega$. The modes for which $z_\mu(\mathbf{k})$ tends to zero at long wavelength are precisely the slow modes identified in (30). So we may approximate (40) by neglecting the fast kinetic modes in the sum over $\mu$. It is then consistent to make use of the approximation $(e^x - 1)^{-1} + \frac{1}{2} \approx x^{-1} + \mathcal{O}(x)$ for $x \ll 1$ in the evaluation of $\mathcal{D}_\mu$; with (24), we obtain

$$\mathcal{D}_\mu = \frac{1}{i\omega - \left(ikz_\mu^{(1)} + (ik)^2 z_\mu^{(2)}\right)} \left(1 + \mathcal{O}(k^2)\right). \tag{41}$$

The final step is the evaluation of $\mathcal{N}_\mu$ in terms of the $k$-expansion of $\langle\phi_\mu|$ and $|\psi_\mu\rangle$; this is accomplished by expressing $\langle\phi_\mu^{(p)}|$ in terms of $|\psi_\mu^{(q)}\rangle$, using (24) in (36), expanding $m_\mu$ in powers of $k$, and identifying the successive orders. To orders $\mathcal{O}(k^0)$ and $\mathcal{O}(k^1)$ respectively, we find

$$\langle\phi_\mu^{(0)}| = \frac{1}{\langle\psi_\mu^{(0)}|\psi_\mu^{(0)}\rangle}\langle\psi_\mu^{(0)}|\,, \tag{42}$$

$$\langle\phi_\mu^{(1)}| = \frac{1}{\langle\psi_\mu^{(0)}|\psi_\mu^{(0)}\rangle}\,\langle\psi_\mu^{(0)}|c_\ell + \langle\psi_\mu^{(1)}| \tag{43}$$

$$- \frac{1}{\langle\psi_\mu^{(0)}|\psi_\mu^{(0)}\rangle^2}\left(\langle\psi_\mu^{(0)}|c_\ell|\psi_\mu^{(0)}\rangle + 2\langle\psi_\mu^{(0)}|\psi_\mu^{(1)}\rangle\right)\langle\psi_\mu^{(0)}|\,, \tag{44}$$

which results are inserted into (40) to yield

$$\mathcal{N}_\mu = \frac{\langle R|\psi_\mu^{(0)}\rangle^2}{\langle\psi_\mu^{(0)}|\psi_\mu^{(0)}\rangle}\Bigg(1 + 2\,ik\,\frac{\langle R|\psi_\mu^{(1)}\rangle}{\langle R|\psi_\mu^{(0)}\rangle} - 2\,ik\,\frac{\langle\psi_\mu^{(0)}|\psi_\mu^{(1)}\rangle}{\langle\psi_\mu^{(0)}|\psi_\mu^{(0)}\rangle} + ik\,\frac{\langle R|c_\ell|\psi_\mu^{(0)}\rangle}{\langle R|\psi_\mu^{(0)}\rangle} - ik\,\frac{\langle\psi_\mu^{(0)}|c_\ell|\psi_\mu^{(0)}\rangle}{\langle\psi_\mu^{(0)}|\psi_\mu^{(0)}\rangle} + \ldots\Bigg)\,. \tag{45}$$

We now discuss the evaluation of $\mathcal{N}_\mu$ for each hydrodynamic mode separately:

• $\mu = \perp$ . As the vectors $\langle R|$ and $|\psi_\perp^{(0)}\rangle$ are orthogonal, $\mathcal{N}_\perp = 0$, and the shear mode will not show up in the density fluctuations power spectrum.

• $\mu = \pm$. To order $\mathcal{O}(k^0)$, $\mathcal{N}_\pm = \frac{b}{2}\kappa_r^2/(\kappa_r + \kappa_b)$. The computation of the next order requires, in principle, complete knowledge of $|\psi_\mu^{(1)}\rangle$, but in fact this is unnecessary because the terms including the unknown $b_{\mu\mu}$ cancel each other; thus we obtain $\mathcal{N}_\pm = \frac{b}{2}\kappa_r^2/(\kappa_r + \kappa_b)(1 \mp ik\Gamma/(2c_s) + \ldots)$.

• $\mu =$ *diff*. To order $\mathcal{O}(k^0)$, $\mathcal{N}_{\text{diff}} = b\kappa_r\kappa_b/(\kappa_r + \kappa_b)$. The contributions to order $\mathcal{O}(k^1)$ all vanish identically; so the expression to $\mathcal{O}(k^0)$ is correct up to corrections of $\mathcal{O}(k^2)$.

Combining $\mathcal{N}_\mu$ with the explicit expression of $\mathcal{D}_\mu$ (using (30), (31) and (34) in (41)), we obtain $F^{red}(\mathbf{k},\omega)$, and therefrom the analytical expression of the dynamic structure factor which reads

$$\rho^{red}\,S^{red}(\mathbf{k},\omega) = b\,\frac{\kappa_r^2}{\kappa_r + \kappa_b}\sum_\pm\left(\frac{\Gamma k^2}{(\omega \pm c_s k)^2 + (\Gamma k^2)^2} + \frac{\Gamma k}{c_s}\frac{c_s k \pm \omega}{(\omega \pm c_s k)^2 + (\Gamma k^2)^2}\right) + b\,\frac{\kappa_r\kappa_b}{\kappa_r + \kappa_b}\frac{2Dk^2}{\omega^2 + (Dk^2)^2}\,. \tag{46}$$

At fixed value of $k$, the spectrum consists of a Brillouin-doublet centered around $\pm kc_s$, and of a central peak characterizing color diffusion. To first order in $k$, $\mathcal{I}m(z_\mu)$ yields the frequency shift of the spectral peak corresponding to the propagating modes ($\mu = \pm$), and to second order in $k$, $\mathcal{R}e(z_\mu(k))$ yields the dissipation coefficients which determine the line-widths of the spectral components. Note that (46) and (11) yield $\int_{-\infty}^{+\infty} d\omega S^{red}(\mathbf{k},\omega)/2\pi S^{red}(\mathbf{k}) = 1$; the spectra shown in the figures are normalized accordingly.

Considering the fluctuations

$$\delta\rho_{\text{diff}}(\mathbf{r},t) \;=\; \rho_{\text{diff}}(\mathbf{r},t) - <\rho_{\text{diff}}(\mathbf{r},t)> \tag{47}$$

of the observable $\rho_{\text{diff}}$ defined as

$$\rho_{\text{diff}} = \frac{\kappa_b}{\kappa_r + \kappa_b}\rho^{red} \;-\; \frac{\kappa_r}{\kappa_r + \kappa_b}\rho^{blue}\,, \tag{48}$$

the corresponding spectral density

$$\rho_{\rm diff}\, S^{diff}(\mathbf{k},\omega) \,=\, \sum_{\mathbf{r}\in\mathcal{L}} \sum_{t=-\infty}^{\infty} e^{-i\omega t - i\mathbf{k}\cdot\mathbf{r}} < \delta\rho_{\rm diff}(\mathbf{r},t)\; \delta\rho_{\rm diff}(0,0) > , \tag{49}$$

can be computed straightforwardly along the lines of the evaluation of the power-spectrum $S^{red}(\mathbf{k},\omega)$; since $|\psi^{(0)}_{\rm diff}\rangle$ is orthogonal to the other eigenvectors, there is no coupling with the other modes, and the dynamic structure factor is a single Lorentzian

$$\rho_{\rm diff}\, S^{diff}(\mathbf{k},\omega) = b\, \frac{\kappa_r \kappa_b}{\kappa_r + \kappa_b}\, \frac{2Dk^2}{\omega^2 + (Dk^2)^2} \,, \tag{50}$$

with similar normalization as for $S^{red}(\mathbf{k},\omega)$. The power spectrum (50) characterizes color diffusion alone, a feature which will be used in the analysis of the simulation data in the next section.

## V. THE POWER SPECTRUM

The above results obtained in the hydrodynamic limit are in accordance with the Landau-Placzek theory for continuous fluids [7]. Now the hydrodynamic theory breaks down at short wavelengths, but the Boltzmann theory should remain valid down to $k$ values where the kinetic domain starts. Lattice gas automata are appropriate model systems to investigate quantitatively the various regimes covering a wide range of wavelengths. In order to characterize the wavelength domain of the various regimes, we consider the quantities $k\ell_f$, where $\ell_f(\sim 1/\rho)$ is the mean free path, and $f = \rho/\rho_{max}$, the reduced density (or the average density per channel). Accordingly the hydrodynamic regime is defined by $k\ell_f \ll 1$, the generalized hydrodynamic regime (Boltzmann regime) by $k\ell_f < 1$, and the kinetic regime by $k\ell_f \gtrsim 1$.

We now discuss the power spectra obtained from automaton simulation data and we compare the results to the analytical Landau-Placzek expressions and to the predictions of the lattice Boltzmann theory. For the latter, we use the eigenvalue spectrum of the propagator $\Gamma$ which can be evaluated numerically over the complete $k$-domain, so extending the computation of the power spectrum to the region of $k$ values where analytical evaluation can no longer be performed. In Fig.1 we show a typical eigenvalue spectrum as computed numerically, where the fourteen modes of the 14-bit model described in Sec.II can be distinguished.

The numerical experiments are performed using a 'color' FHP-3 lattice gas (see Sec.II) at equilibrium. The lattice size is 256 $\times$ 256 nodes, and the simulation duration is 40.000 time steps. Spatial Fourier transforms are computed at every time-step, and time-Fourier transforms are taken over intervals of 16384 time-steps shifted by 20 time-steps for averaging; data shown in the figures are smoothed by low-pass frequency filtering.

In Fig.1.a we observe that in the range $0 < k < 0.4$, corresponding to wavelengths $\lambda > 15\ell_0$ (with $\ell_0$ the lattice unit length), the four slow modes are well separated from the kinetic modes, and their behavior is correctly given by the hydrodynamic expressions. We therefore expect that in this domain, the Landau-Placzek theory should provide a correct description of the dynamic structure factor. This is indeed confirmed by the comparison between the results obtained from simulation data and the theoretical predictions as shown in Fig.2, where we also notice that the Landau-Placzek spectrum is indistinguishable from the Boltzmann spectrum.

The diffusion coefficient $D(f,\theta)$ is obtained straightforwardly from the diffusive mode power spectrum: $S^{diff}(\mathbf{k},\omega)$ is a single Lorentzian (50) with half-width $\Delta\omega = Dk^2$; so plotting $\Delta\omega$ as a function of $k^2$ (see Fig.3.a) yields, by least-squares fit, a slope whose value provides an experimental measure of $D$. In Figure 3.b, we show the diffusion coefficient as a function of density: we observe that the exact Boltzmann result [15] is in good agreement with the lattice gas simulation data up to $f \approx 0.25$; for larger $f$ the theoretical prediction deviates progressively from the measured values indicating that the molecular chaos assumption becomes invalid at high densities.

As $k$ increases from 0.4 to 1.4, there is still a distinct scale separation between slow and fast modes (see Fig.1), but the eigenvalues of the slow modes depart significantly from the hydrodynamic prediction, indicating the breakdown of the local response hypothesis: the transport coefficients become $k$-dependent. As a result, the Landau-Placzek theory does no longer describe the power spectrum correctly – for instance, there is a noticable spectral line broadening in $S^{diff}(k,\omega)$ (see Fig.4.a) –, but the complete Boltzmann spectrum is in good agreement with the simulation results, as seen in Figs.4.a and 4.b. We have also observed that even at rather short wavelength ($\lambda \sim 10\ell_0$) the experimental data can still be approximated with a Landau-Placzek spectral function if the transport coefficients are parameterized, showing that a hydrodynamic type description holds qualitatively down to quite short wavelengths. For $k > 1.5$, all modes exhibit comparable decay rates (see Fig.1.a), and all modes with even parity in $c_\perp$ contribute significantly to the power spectrum. In this domain, the Landau-Placzek theory is invalid and the Boltzmann computation provides good agreement with the experimental spectrum (down to $\lambda \approx 4\ell_0$) as shown in Fig.4.b.

From the agreement between the experimental data and the Boltzmann results, we found that the Boltzmann theory remains valid up to reduced densities of $f \approx 0.25$. At higher densities the discrepancy between the Boltzmann spectral density and the experimental power spectrum (see Fig.5.a) reflects the failure of the Boltzmann theory to evaluate correctly the transport coefficients. Here the contributions of 'ring-collisions' [16] should be included in the evaluation of the diffusion coefficient. Finally we note that at *very* high densities ($f = .9$), we observe a slight coupling between the color diffusion mode and the sound propagation modes as illustrated in Fig.5.b; so far we have no theoretical analysis of this effect.

We close this section with some comments on the spurious invariants which have been known to plague lattice gas automata with spurious conservation laws [17]. The present model is not free of this 'contamination': figure 6.a shows clearly that $\mathcal{R}e(z_\perp(k)) \to 0$ (i.e. $|e^{z_\perp(k)}| \to 1$) as $k \to 2\pi$ (which corresponds to a wavelength of one lattice-unit), denoting a mode which persists once it is excited. The corresponding spurious invariance can be interpreted as the conservation of total transverse momentum on even and odd lines of the lattice every two time-steps ($\mathcal{I}m(z_\perp(k)) \to \pi$, and therefore $e^{z_\perp(k)} \to -1$ ). Consequently it is important to choose initial conditions such that the total transverse momentum on odd and even lines be both rigorously zero. This may easily be realized by implementing the particles pairwise with opposite velocities on the same node. Without this precaution, the power spectrum may exhibit spurious effects as illustrated in Fig.6.b. The full line shows the correct spectrum. Now if we denote the reference direction by $x$, and measure the spectrum with $\mathbf{k}$ oriented along the $y$-direction (orthogonal to the $x$-axis), we obtain the spectrum represented by the dotted line in Fig.6.b: the speed of sound is not affected, but the line-width of the shifted peaks is incorrect, and the diffusive mode is propagative, despite the fact that the spectrum is measured in the long wave-length domain ( $\lambda \approx 64\ell_0$). With $\mathbf{k}$ oriented at an angle of 30 deg. off the $x$-axis, we observe (see dot-dashed spectrum in Fig.6.b) that there is a significant shift in the speed of sound, that the diffusive mode is practically absent, and that the kinetic modes which couple to the transverse momentum invade the spectrum.

## VI. COMMENTS

We have presented a two species non-thermal lattice gas automaton for which we have developed the lattice Boltzmann theory and performed automaton simulations. The emphasis is on the fluctuation correlations in order to obtain a microscopic analysis of diffusion dynamics in contrast to earlier studies based on macroscopic approaches. On large space- and time-scales we find spectral features of the dynamic structure factor in accordance with those of real fluids described by the Landau-Placzek theory. Because of the intrinsic simplicity of the lattice gas model, the various wavelength regimes can be easily identified, and the propagator spectrum can be used to to compute the power spectrum over the full wavenumber domain, and to test the the validity of the Boltzmann hypothesis. The present study based on the analysis of spontaneous fluctuations offers a microscopic approach to diffusion , and, through the identification of a purely diffusive mode associated to color transport, complements and supports earlier macroscopic investigations.

### Acknowledgments

We thank Alberto Suárez, Olivier Tribel and Jörg Weimar for helpful discussions during the course of this work. DH has benefited from a grant from the *Fonds pour la formation à la Recherche dans l'Industrie et l'Agriculture* (FRIA, Belgium). JPB acknowledges support by the *Fonds National de la Recherche Scientifique* (FNRS, Belgium).

---

FIG. 1. Eigenvalue spectrum of the 14-bit model propagator: Boltzmann computation (full lines) and hydrodynamic limit (dashed lines). The reduced density and concentration are $f = 0.15$ and $\theta_{red} = 30\%$ respectively. The wave-number $k = |\mathbf{k}|$ is given in reciprocal lattice units

FIG. 2. Spectral measurement of the diffusion coefficient: (a) $\Delta\omega = Dk^2$, $f = 0.3$ and $\theta_{red} = 30\%$ (open squares), 50% (black dots); the least-squares fits (solid line) coincide. (b) $D = D(f, \theta_{red})$, simulation data (black dots) and Boltzmann prediction (solid curve); the size of the black dots corresponds to the largest error bar ($|\Delta D/D| \leq 2\%$).

FIG. 3. Power spectra of 'red' density fluctuations (a) and of $\rho_{\rm diff}$ fluctuations (b) at low density and small $k$. Comparison of experimental data (full lines) with theoretical predictions: the Boltzmann results and the Landau-Placzek spectra coincide (dashed lines). Density $f = 0.15$; concentration $\theta_{red} = 30\%$; wave number $|\mathbf{k}| = 0.098$ reciprocal lattice units; $\omega$ is given in reciprocal time units ($2\pi/T$, where $T$ is total number of time-steps); the spectral functions are given in reciprocal $\omega$ units.

FIG. 4. Power spectra at low density and high $k$: $|\mathbf{k}| = 0.49$ (a), i.e. $\lambda \approx 12\ell_0$; $|\mathbf{k}| = 1.57$ (b), i.e. $\lambda \approx 4\ell_0$. Experimental data (full lines), Boltzmann spectrum (dashed lines), Landau-Placzek theory (doted lines). $f = 0.15$; $\theta_{red} = 30\%$.

FIG. 5. Spectra at high density: $f = 0.5$ (a) and $f = 0.9$ (b). Experimental data (full lines) and Boltzmann prediction (dashed line). Concentration $\theta_{red} = 30\%$; wave number $|\mathbf{k}| = 0.098$.

FIG. 6. Effect of spurious invariant. (a) Boltzmann propagator eigenvalues (full lines) with spurious invariant (dashed line); (b) shows the results of two simulations: the spectrum plotted as a full line is obtained with initial conditions for which the oscillating transverse mode is not excited; the two other power spectra are obtained for two $\mathbf{k}$'s with different orientations using 'incorrect' initial conditions (see text). Density $f = 0.10$; concentration $\theta_{red} = 50\%$; wave number $|\mathbf{k}| = 0.098$.

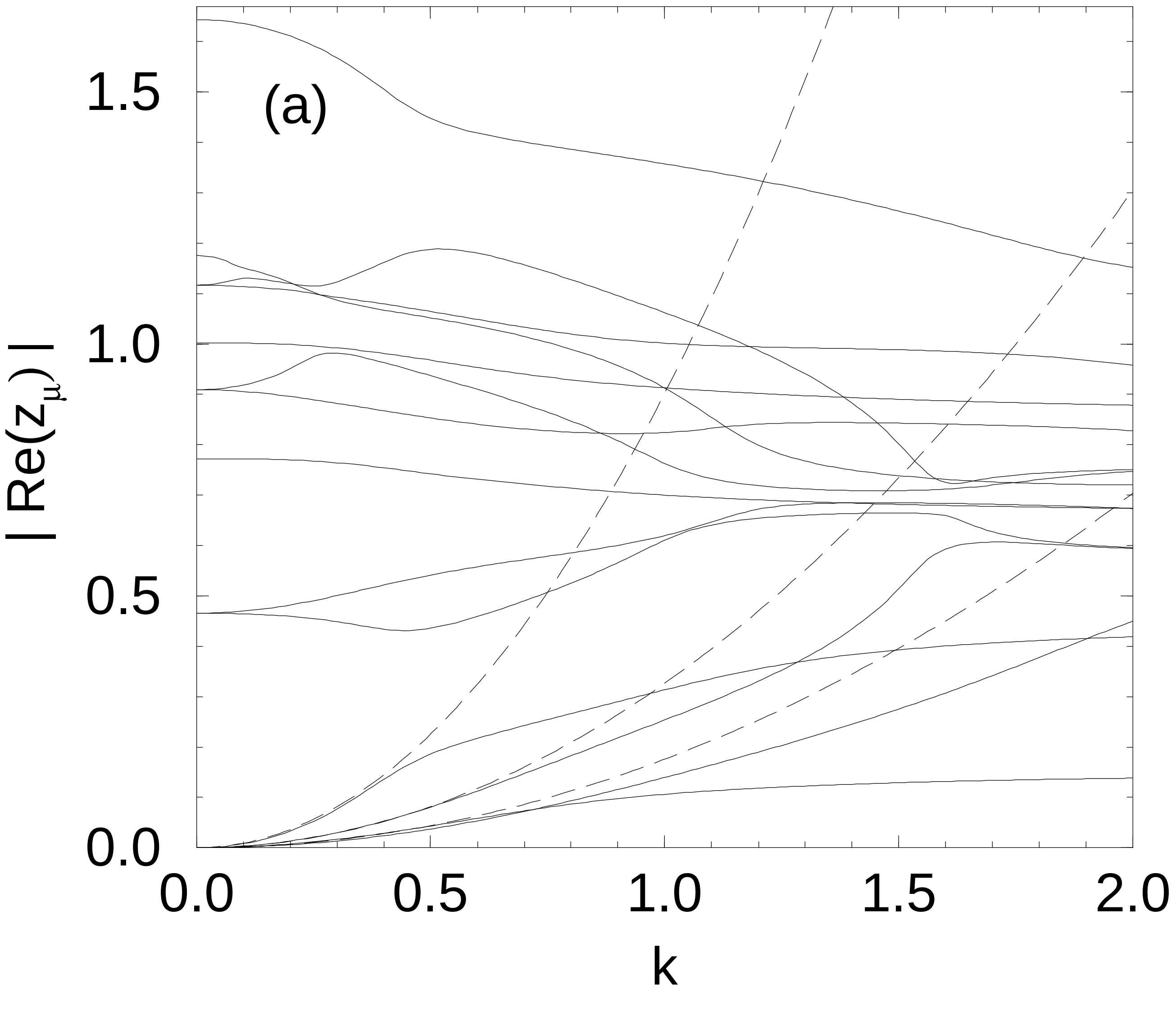
(a)
1.5
1.0
0.5
0.0
0.0
0.5
1.0
1.5
2.0
| Re($z_\mu$) |
k

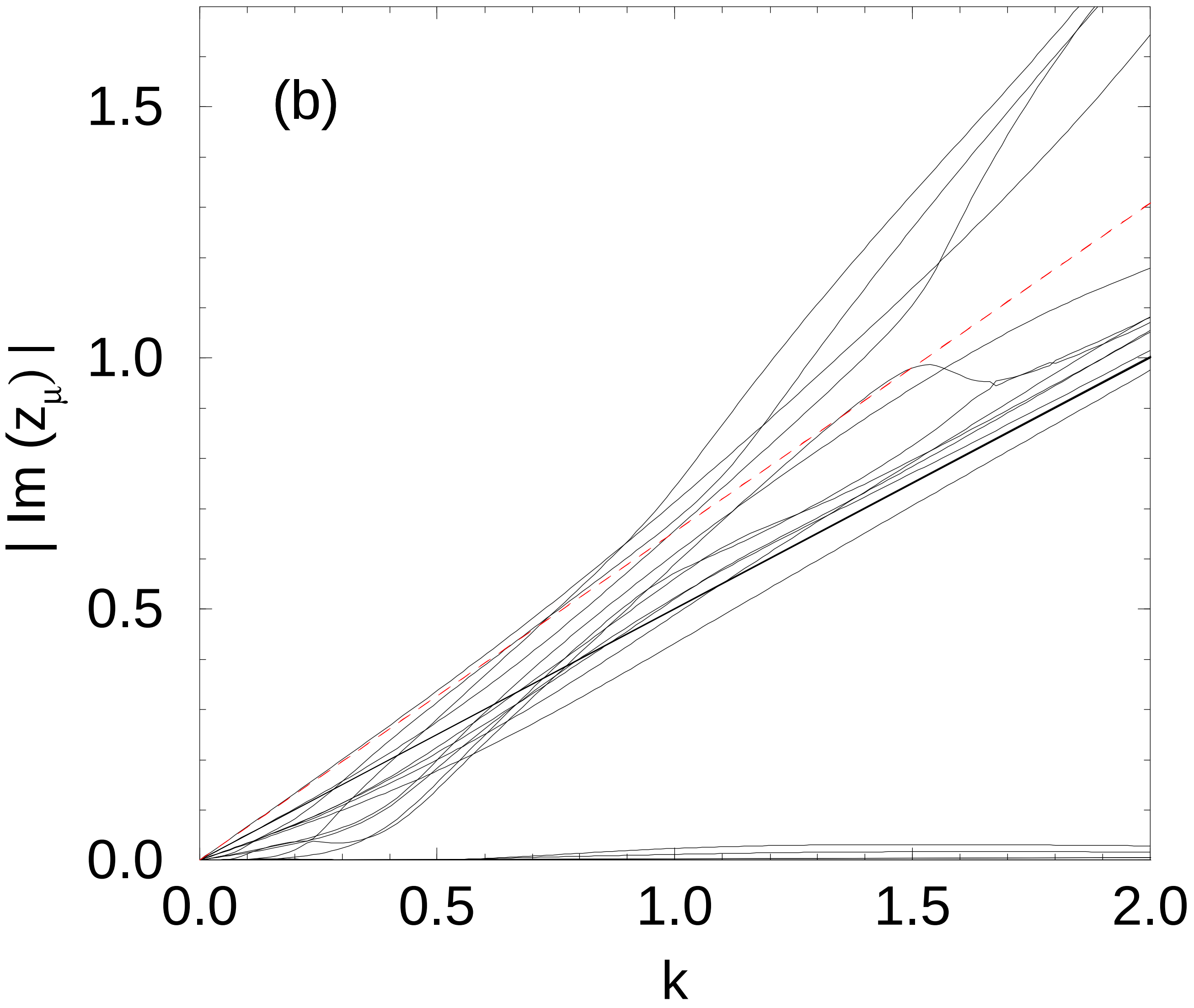

(b)
1.5
1.0
0.5
0.0
| Im (z_μ) |
0.0
0.5
1.0
1.5
2.0
k

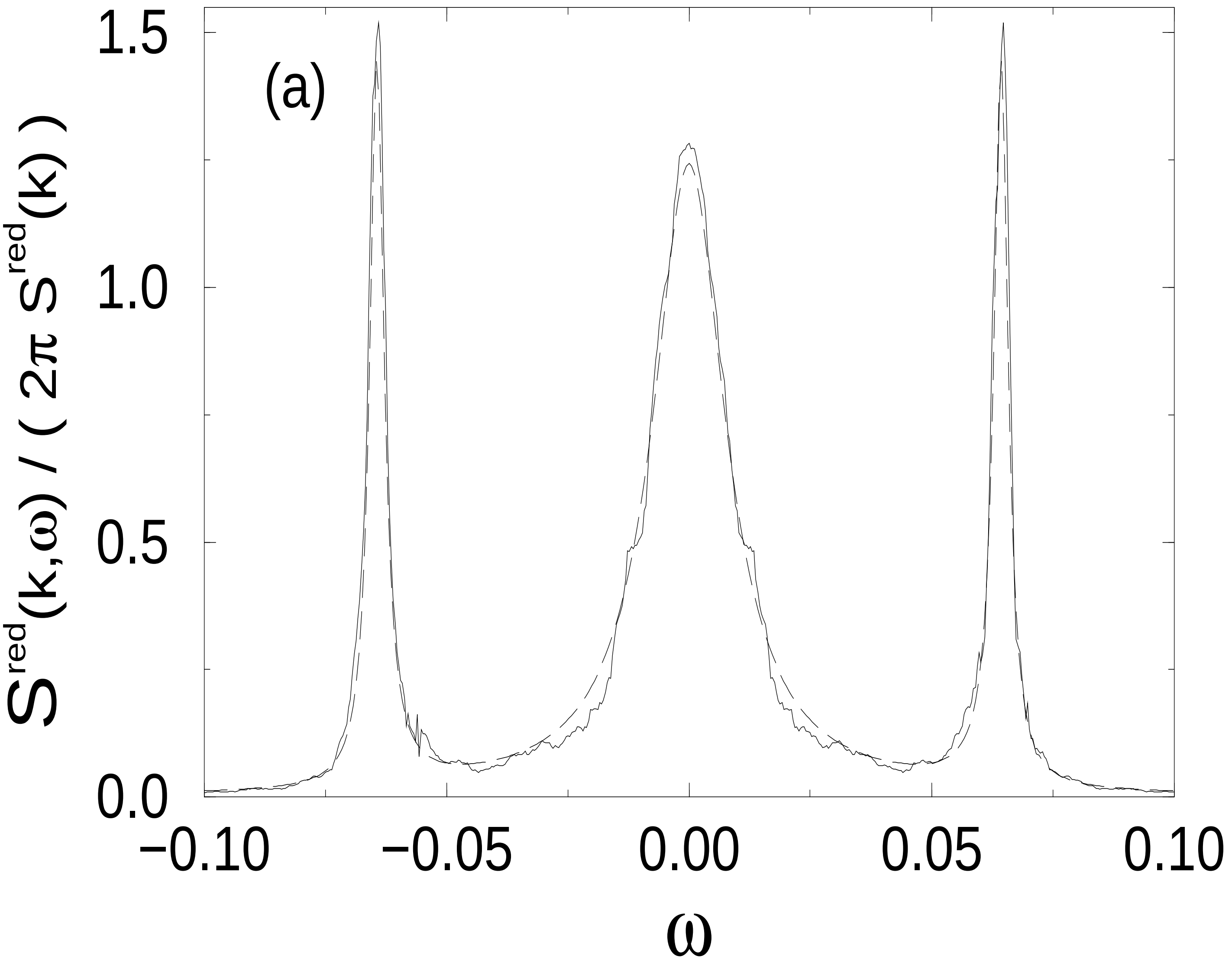

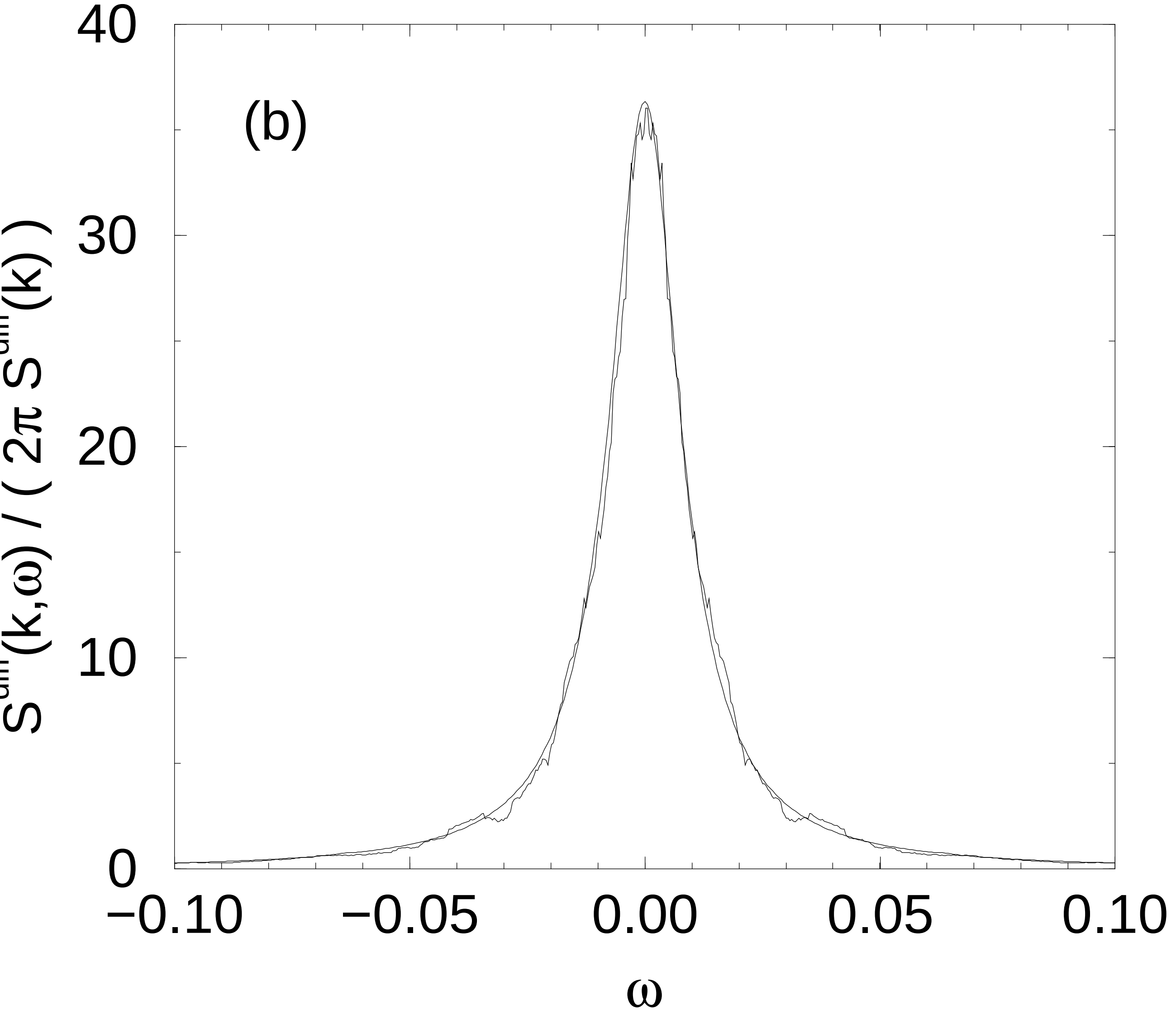

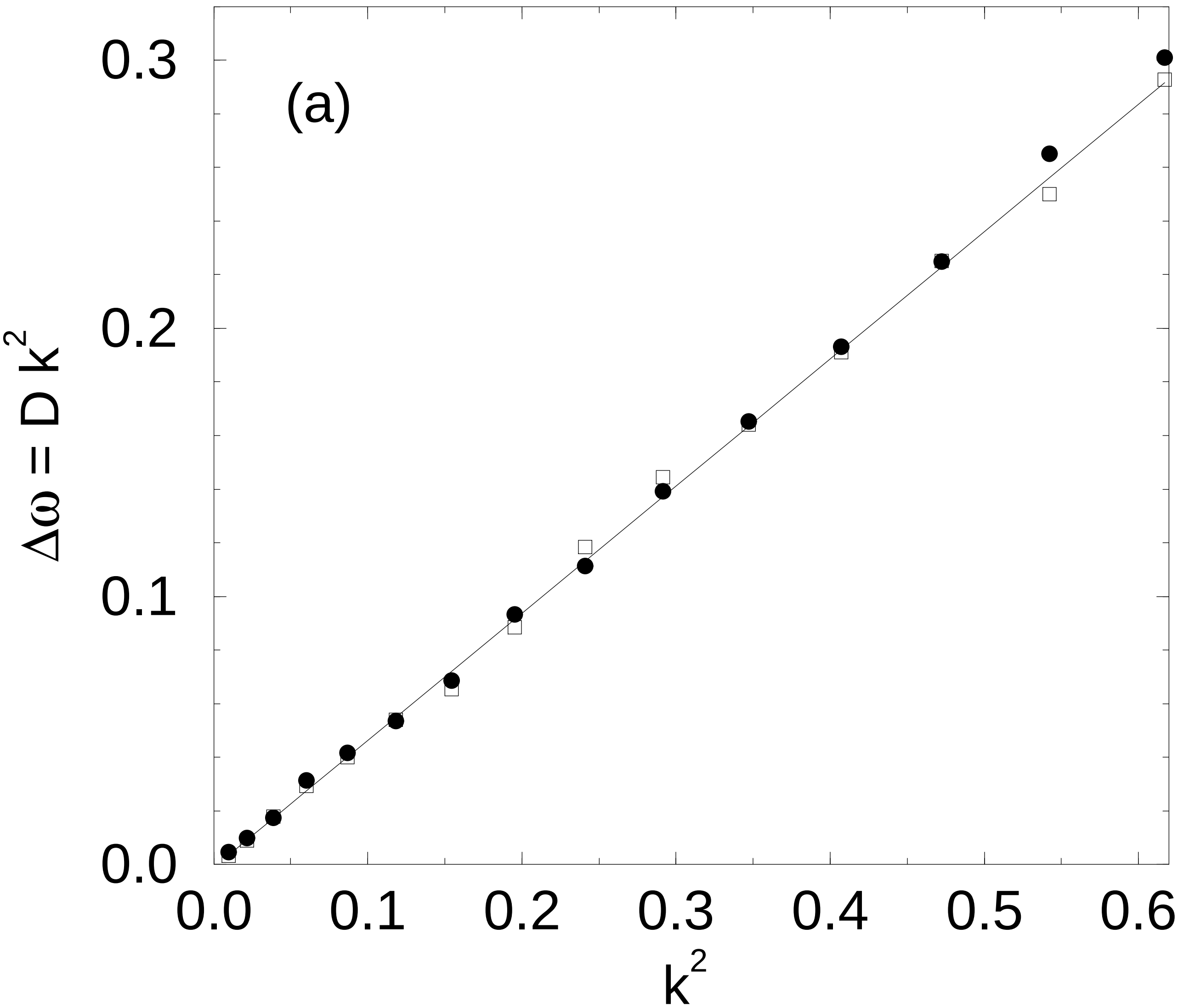

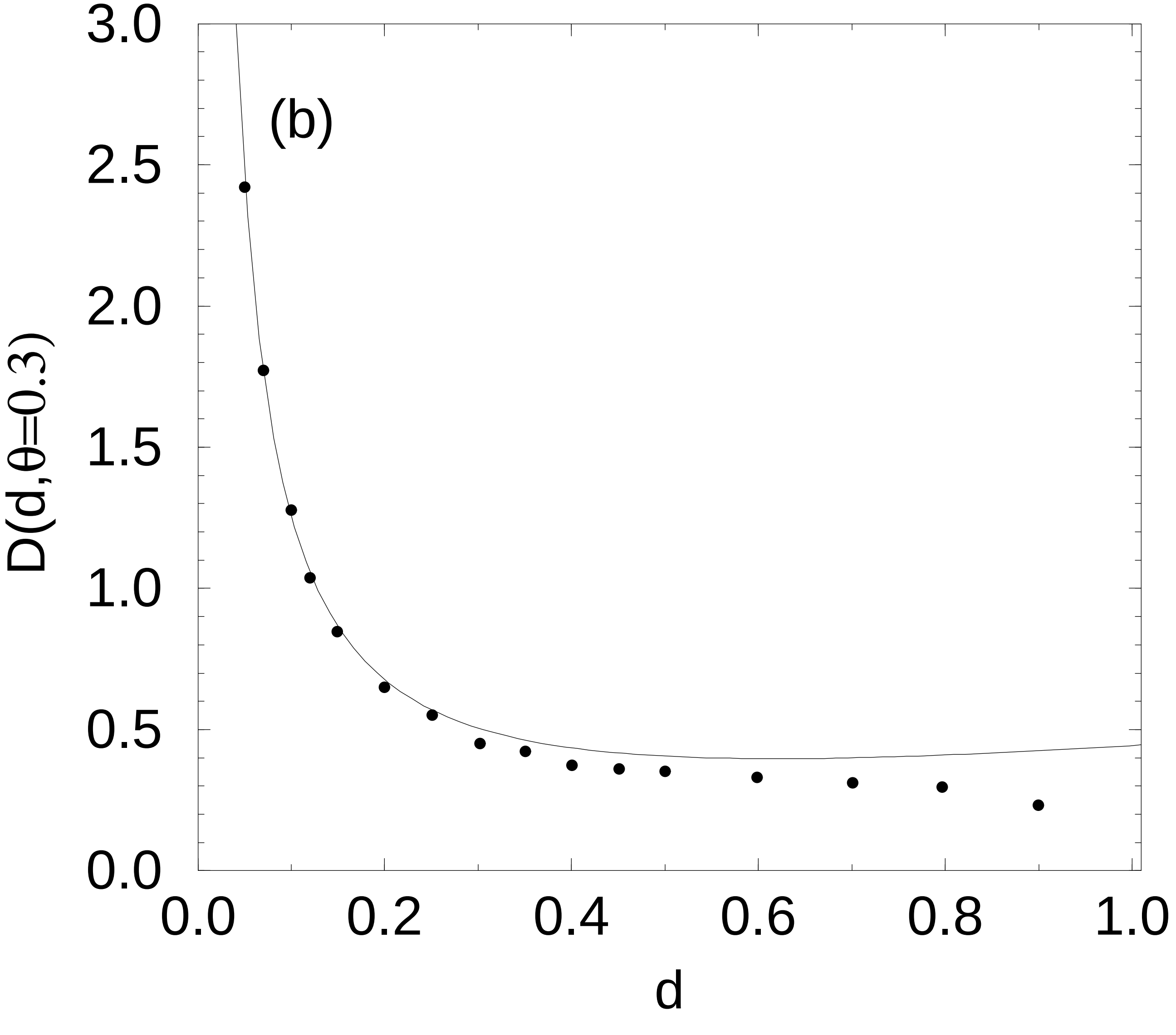

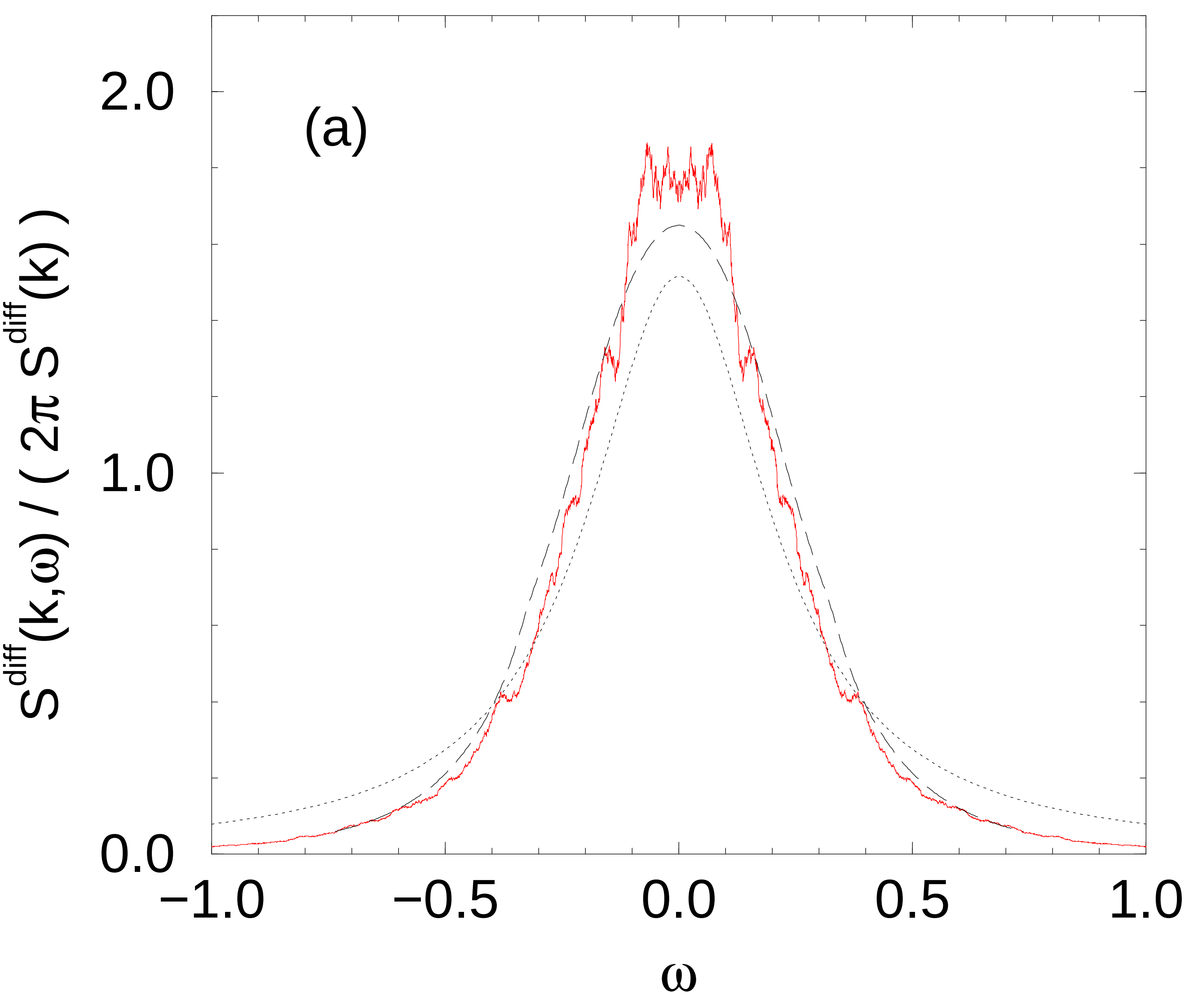

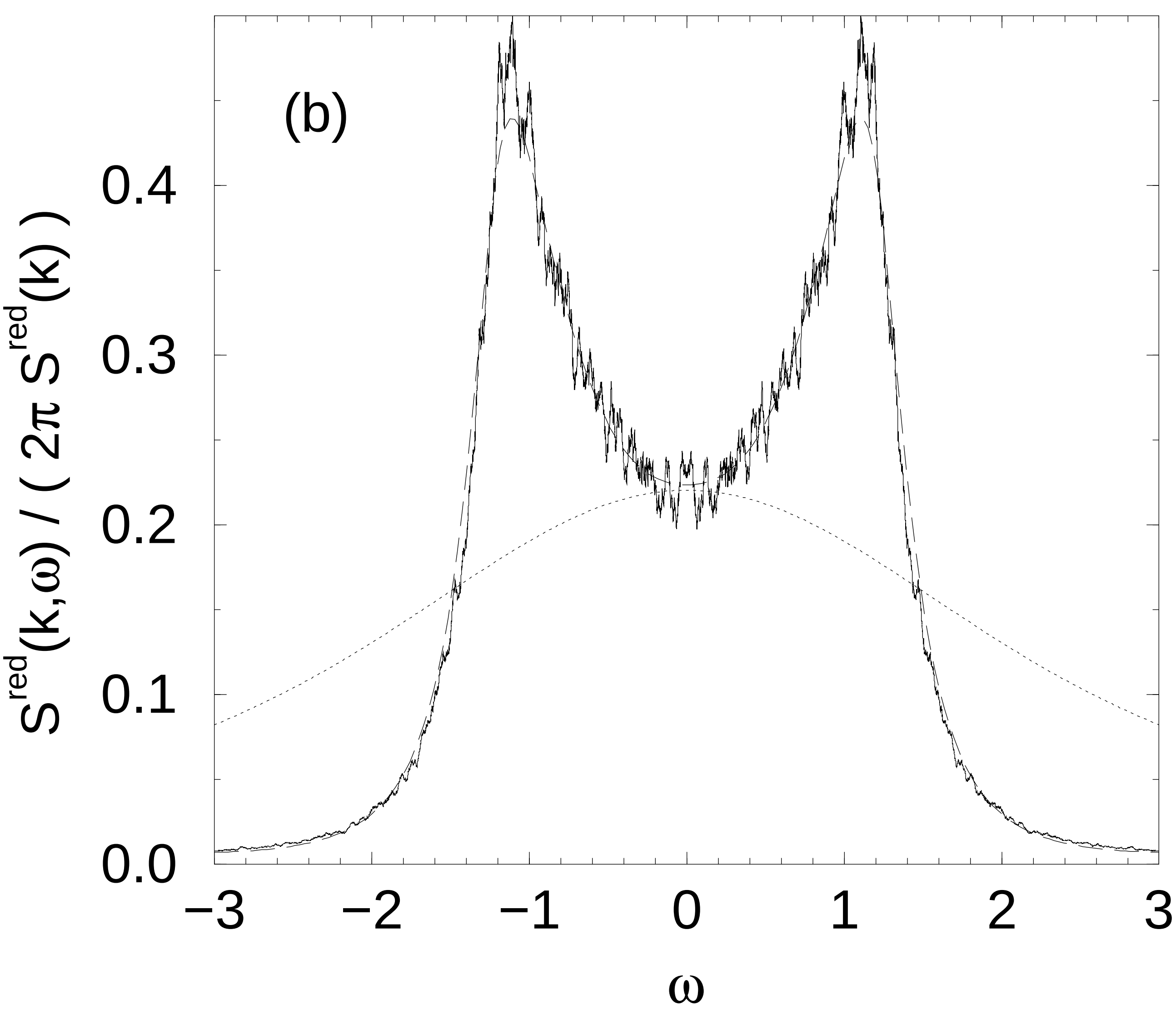

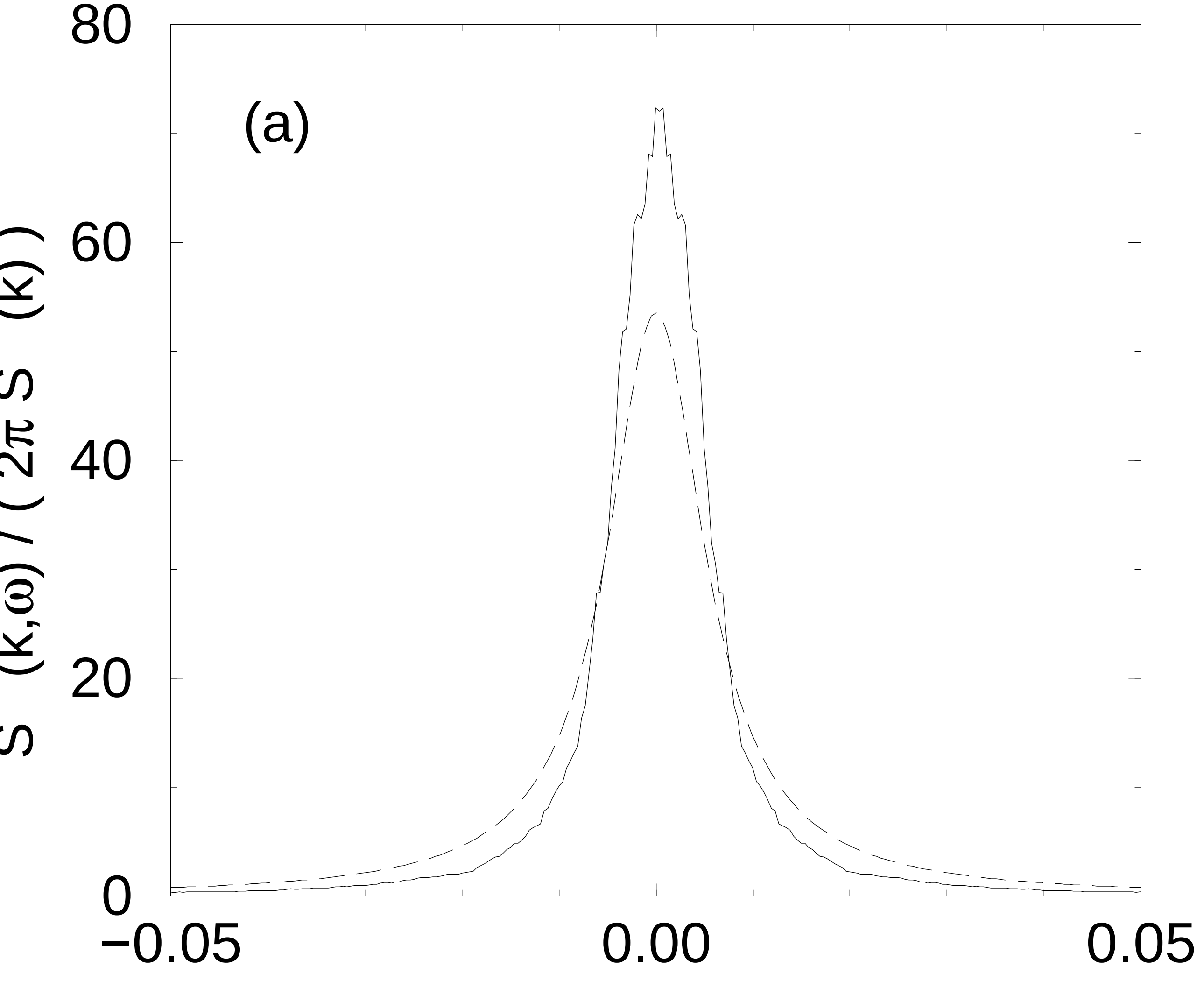

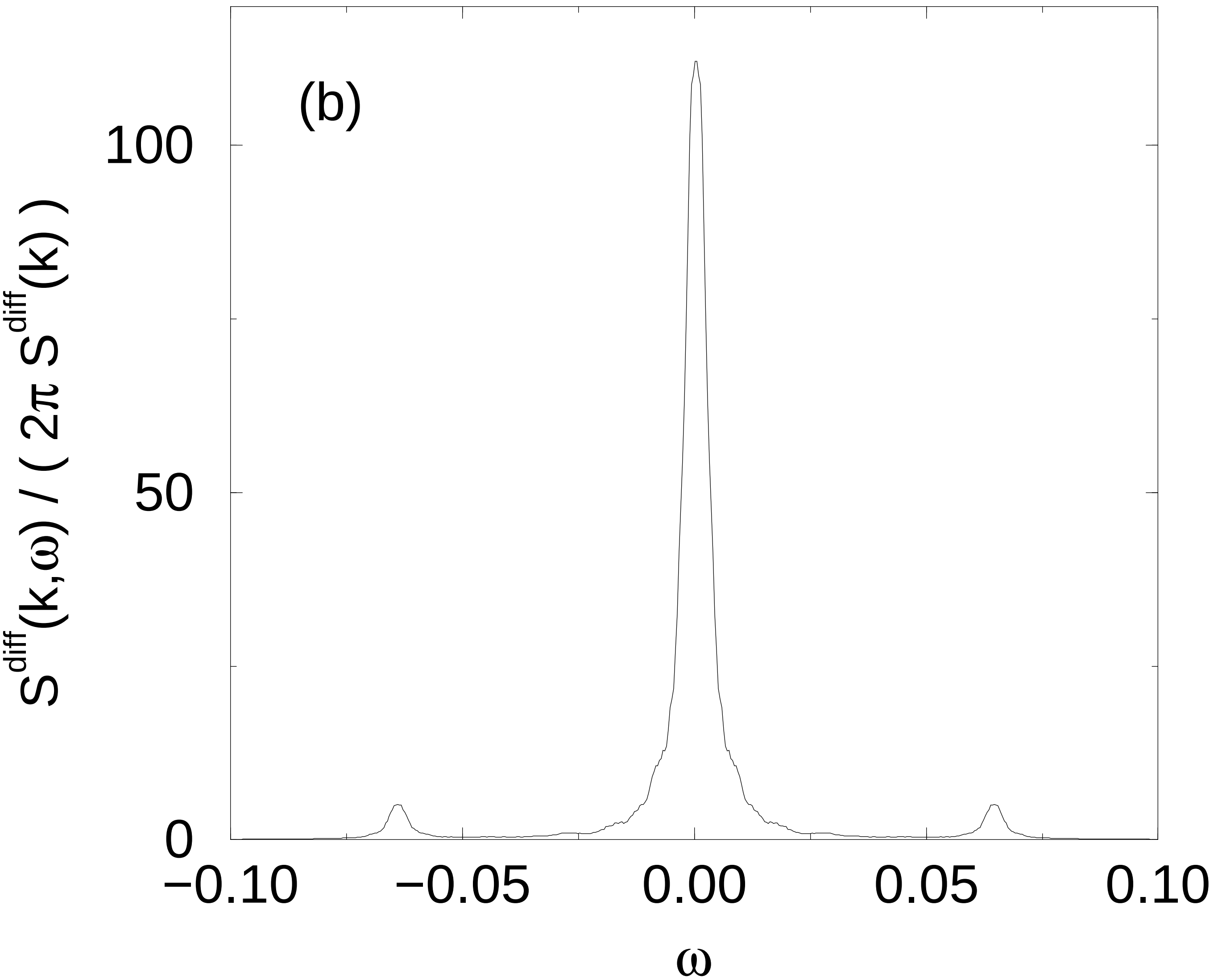

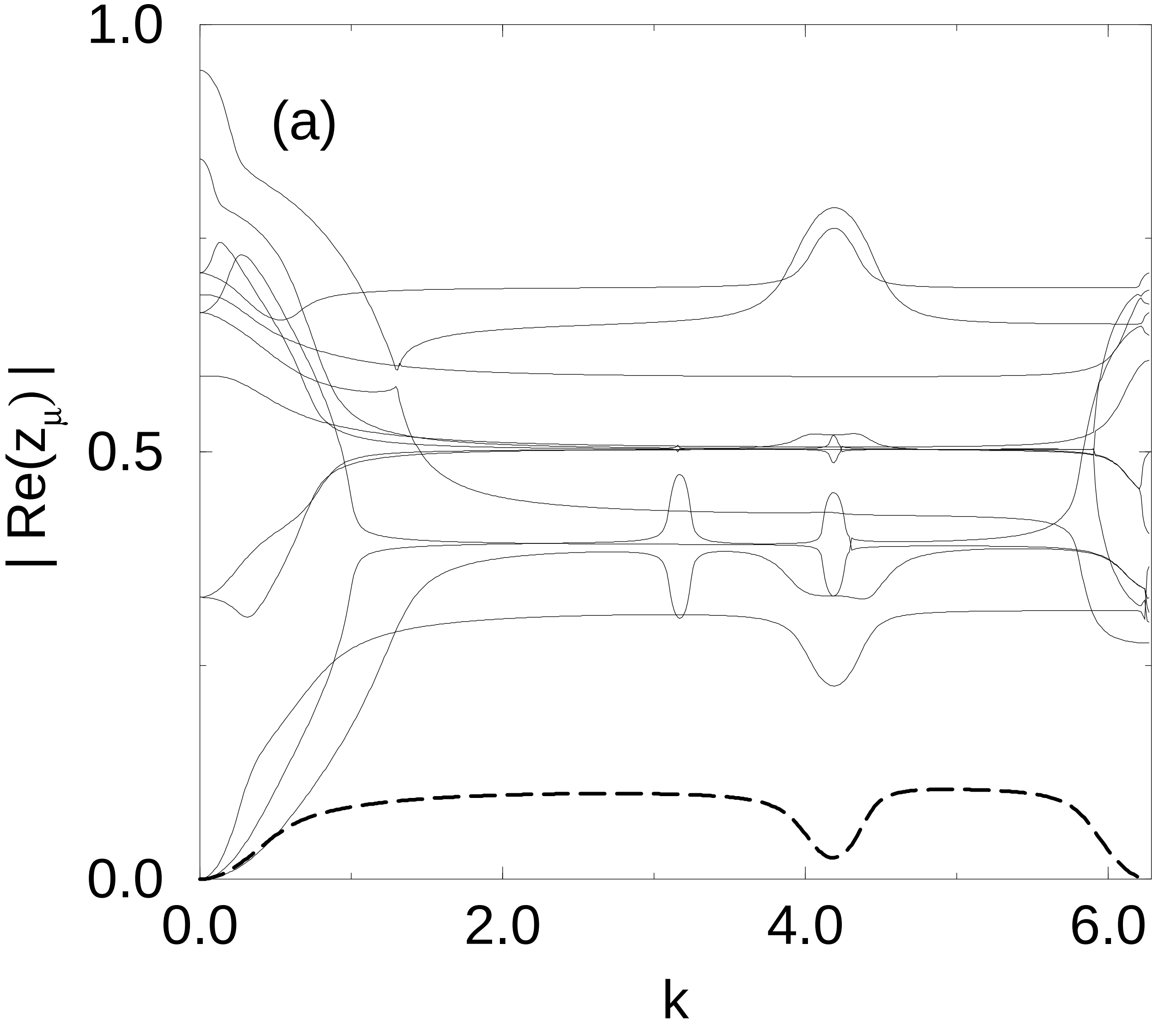
1.0
(a)
0.5
0.0
|Re(z_μ)|
0.0
2.0
4.0
6.0
k

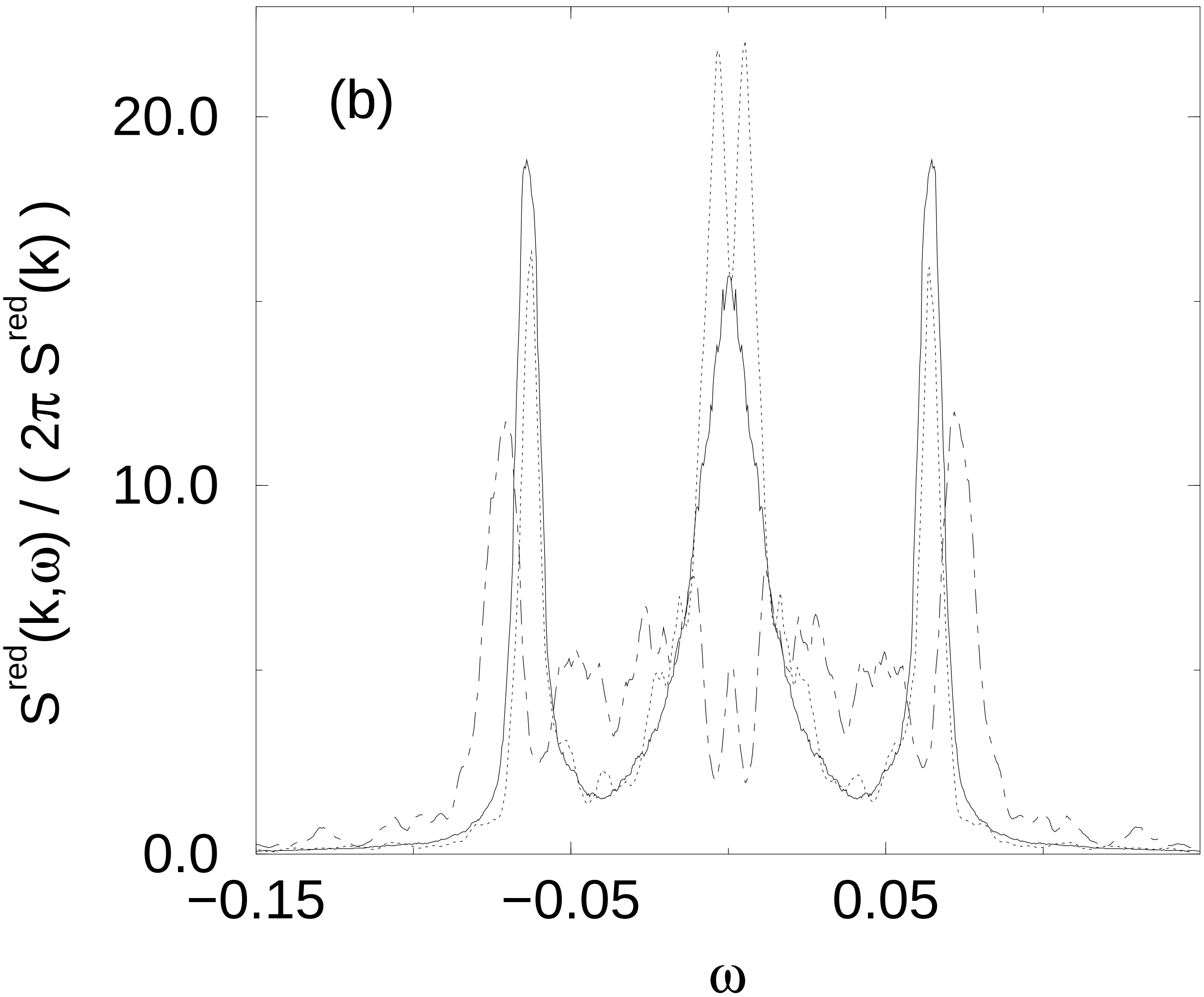